# Magnetic form factor of SrFe$_2$As$_2$

William Ratcliff II[1*], P. A. Kienzle[1], Jeffrey W. Lynn[1], Shiliang Li[2,4], Pengcheng Dai[2,3,4], G. F. Chen[4], and N. L. Wang[4]

[1]NIST Center for Neutron Research, National Institute of Standards and Technology, Gaithersburg, Maryland 20899, USA
[2]Department of Physics and Astronomy, University of Tennessee, Knoxville, TN 37996
[3]Neutron Scattering Science Division, Oak Ridge National Laboratory, Oak Ridge, TN 37831
[4]Beijing National Laboratory for Condensed Matter Physics, Institute of Physics, Chinese Academy of Sciences, Beijing 100080, People's Republic of China

**Abstract**

Neutron diffraction measurements have been carried out to investigate the magnetic form factor of the parent SrFe$_2$As$_2$ system of the iron-based superconductors. The general feature is that the form factor is approximately isotropic in wave vector, indicating that multiple *d*-orbitals of the iron atoms are occupied as expected based on band theory. Inversion of the diffraction data suggests that there is some elongation of the spin density toward the As atoms. We have also extended the diffraction measurements to investigate a possible jump in the *c*-axis lattice parameter at the structural phase transition, but find no detectable change within the experimental uncertainties.

## I. Introduction

The nature of the magnetic moment and the spin configuration of the long range antiferromagnetic order in the iron-based pnictide superconductor class of materials is a topic of great current interest. Antiferromagnetic order develops at or just below a structural distortion that breaks the tetragonal symmetry [1]. In the distorted *a-b* plane the spins align antiparallel along the longer *a*-axis while they align parallel along the *b*-axis. Hence the distortion plays an essential role in the magnetic structure since this magnetic configuration does not have tetragonal symmetry, and a number of theories suggest that the mechanism of the structural distortion is magnetic [2-4]. We have carried out quantitative measurements of the magnetic Bragg intensities for SrFe$_2$As$_2$ to determine the magnetic form factor, which is directly related to the magnetization density in the unit cell of the crystal through Fourier inversion. We find that the dominant spin density resides on the iron as expected, and is approximately isotropic indicating that multiple *d* orbitals are occupied [4]. This approximate isotropy strongly contrasts with other S=½ magnets such as for the undoped cuprate superconductors systems [5] as well as K$_2$IrCl$_6$ [6], where the form factor is highly anisotropic and has orbital bonding character. In the present system there also is some modest anisotropy, and this appears to originate from iron-arsenic bonding.

## II. Experimental Procedures

Neutron diffraction measurements were carried out to study the structural transition and magnetic order in this material. Data were collected on the BT-9 triple-axis spectrometer at the NIST Center for Neutron Research. The neutron energies were fixed through the use of pyrolytic graphite (PG) (002) monochromator and analyzer. Measurements of the magnetic form factor were carried out at 14.7 meV, and also with 35 meV to access more reflections and to evaluate the possible role of extinction, which was found not to be a problem. Relatively relaxed Söller collimations of 40′-23′-S-40′-120′ full-width-at-half maximum (FWHM) were employed. To determine the variation of the *c*-axis lattice parameter through the structural phase transition, tight Söller collimations of 10′-10′-S-10′-80′ FWHM were utilized at a neutron energy of 14.7 meV. PG filters were placed both before and after the sample to suppress higher order wavelengths to negligible levels. The single crystal measured was the same one used in a previous study, with orthorhombic lattice parameters $a \approx b \approx 5.57$ Å and $c \approx 12.29$ Å [7]. Magnetic reflections were measured in the [H, 0, L] zone in a helium cryostat. Uncertainties where indicated are statistical in origin and represent one standard deviation.

### III. Form Factor Results and Discussion

The canonical equation for the differential cross section describing the coherent elastic scattering of neutrons from magnetically ordered crystals in the ground state is given by: [6,8,9]

$$\frac{d\sigma}{d\omega} \propto N_M \frac{2\pi^3}{V_M} \sum_{\vec{G}_M} \delta(\vec{Q} - \vec{G}_M) | \vec{F_M}(\vec{G}_M) |^2 \quad (1)$$

where $N_M$ is the number of magnetic unit cells in the crystal and $V_M$ is the volume of the magnetic unit cell. For a simple collinear magnetic structure the vector magnetic structure factor $\vec{F}_M$ is related to the scalar magnetic structure factor by

$$\left| \vec{F_M}(G) \right|^2 = \left( \frac{\gamma e^2}{2mc^2} \right)^2 \left\langle 1 - \left( \hat{G} \cdot \hat{M} \right)^2 \right\rangle \left| F_M(G) \right|^2$$
(2)

where the neutron-electron coupling constant in parenthesis is $-0.27 \times 10^{-12}$ cm, $\hat{G}$ and $\hat{M}$ are unit vectors in the direction of the reciprocal lattice vector $\vec{G}$ and the spin direction, respectively, and the scalar structure factor is

$$F_M(G) = \sum_{j=1}^{N} \langle \mu_j \rangle f_j(G) e^{iG \cdot r_j} e^{-W} . \quad (3)$$

Here $\langle \mu_j \rangle$ is the thermal average of the ordered magnetic moment of the $j^{th}$ atom in the unit cell, $r_j$ is the position of the $j^{th}$ atom in the magnetic unit cell, $f_j(\vec{Q})$ is the scalar magnetic form factor of the $j^{th}$ atom in the cell, and the sum $j$ ranges over all atoms in the unit cell. The (scalar) magnetic form factor is the quantity of direct interest here, and is Fourier transform of the magnetization density associated with each atom.

In an earlier work we determined the magnetic structure and order parameter of $SrFe_2As_2$, which consists of antiparallel Fe spins along the *a* and *c* directions and parallel spins along *b*, with the spin direction along *a* [7]. Here we take this spin arrangement as a starting point and extract the magnetic form factor. The results, based on fits to the integrated intensities measured at base temperature (4 K) of 13 independent magnetic reflections (indexed on the basis of the orthorhombic cell), are shown in Fig. 1. The solid curve represents the tabulated isotropic form factor for metallic Fe [10], and the general overall agreement with the measurements indicates that the magnetization is isotropic to a good approximation. Note in particular that reflections of predominately (0,0,L) character (represented by circles in Fig. 1) and those of (H,0,0) character (represented by squares) both follow a similar curve, indicating that the form factor is approximately isotropic. This behavior is in stark contrast to what was found in other $S=\frac{1}{2}$ systems such as the high $T_C$ cuprate family [5] and $K_2IrCl_6$ [6], where the magnetic form factor originates from a single type of orbital that renders them highly directional.

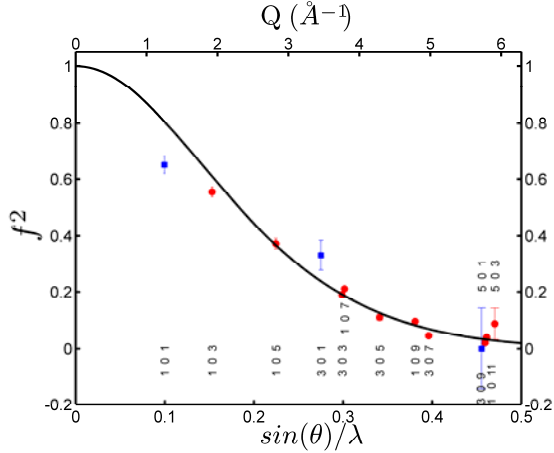

Figure 1. (color online) Measurement of the square of the magnetic form factor as a function of $Q = 4\pi\sin(\theta)/\lambda$. Circles denote reflections which are predominantly along (0,0,L). Squares represent reflections which are predominantly along (H,0,0). The solid curve shows the form factor of metallic iron [4]. Error bars are statistical in nature and represent one standard deviation.

Further perspective of the spatial distribution of the magnetization density can be gained through a real space representation via direct Fourier inversion of the form factor. In measurements of neutron intensities we only determine the magnitude of the structure factor, not its phase. In the present case the structure factors are real since we have a centrosymmetric structure, and we assign the sign of the structure factor based on our previously determined spin structure. The results of the Fourier inversion are shown in Fig. 2(a). As our data were obtained in the (H,0,L) zone, we obtain the projection of the magnetization density along the *b*-axis of the crystal. In the figure, we overlay the density plot with the atomic structure. The two grey (color) levels represent positive and negative magnetizations and show that we recover the antiferromagnetic structure of our original model. However, we notice that there are relatively long tails of the magnetization density which suggest hybridization along the *a*-axis with the As atoms above and below the plane containing the Fe atoms in this projection.

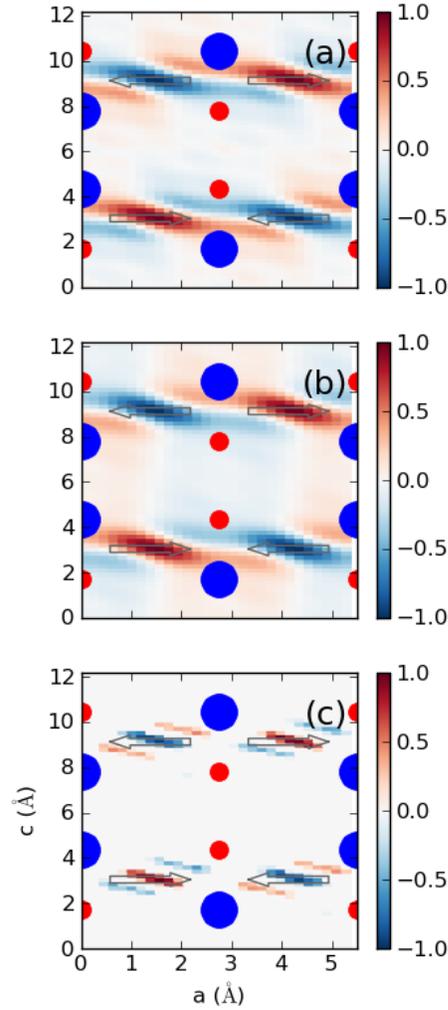

Figure 2. (color online) (a) Fourier inversion of the experimental form factor, yielding the projection of the magnetization density onto the *a-c* plane of the crystal structure. The different (red and blue) shaded regions with the arrows reversed indicate the oppositely directed magnetization of the antiferromagnetic structure. Note the elongation of the density along the Fe-As bond, indicated for the As ions above (small red circles) for one spin direction and below (large blue circles) the Fe plane for the other spin direction. (b) Magnetization density in the Fe plane obtained from the simulated data, where higher Q as well as K≠0 reflections have been calculated and included in the inversion. The similarity of the two plots demonstrates that the features obtained from the actual data are robust against termination effects and the absence of K≠0 data. (c) Maximum entropy reconstruction of the magnetization density, which also yields the same basic magnetization density. For all plots, the densities have been normalized.

One question to address is whether sufficient data have been measured to obtain a reliable magnetization density. To test the reliability of the inversion and possible effects of peaks that were not obtained experimentally, we calculated form factor values, including K≠0 (out-of-scattering-plane) reflections as well as higher Q reflections. The calculated values were based on the known magnetic structure and an assumed spherical Fe form factor. Then the form factor data were Fourier inverted. Figure 2(b) shows a cut of the real space *a-c* plane that contains the Fe, which can be directly compared with the projection in Fig. 2(a). We see that the magnetization densities in the two plots are quite similar, demonstrating that the features obtained from the actual data are robust against termination effects and the absence of K≠0 data.

An alternative to the direct Fourier inversion of the data to obtain the magnetization density is to carry out a maximum entropy reconstruction of the moment density. The basic idea behind maximum entropy is that there may be a number of possible moment densities that fit the form factor data equally well within the experimental uncertainties. Thus to obtain a representative moment density, the strategy is to search for moment densities which maximize entropy, while constrained to minimize the fit to the data. This technique picks the most likely magnetization density consistent with the data. While there are many different algorithms for obtaining the maximum entropy solution, we used a program called ALGENCAN (see [11]) to treat the maximum entropy reconstruction as a constrained optimization problem. The results of the maximum entropy approach are shown in Fig. 2(c). Here, we find the same type of anisotropy in the moment density, although not as pronounced as in the direct Fourier reconstruction. Note that in Fig. 2 the maximum magnetizations have been normalized to be the same in each part, so that in the Fig. 2(c) the magnetization density falls off more quickly than in Fig. (a,b). We remark that the results in Fig. 2c are consistent with those obtained using the PRIMA maximum entropy program [12, 13], and note that both reconstructions suggest that the Fe magnetization density tends to be elongated towards the As atoms. In all reconstructions, we also note that there is a modulation of the moment along the non-elongated direction, which may also be an indication of bonding. It would be particularly interesting to determine if these features can be reproduced theoretically.

In general an electron in a solid has a wave function of the form $u(r)e^{i\mathbf{k}\cdot\mathbf{r}}$, where $u(r)$ must be compatible with the symmetry of the lattice but basically looks like an atomic wave function, *s*, *p*, *d*, etc. It should be noted that the shape of the wave function is unrelated to whether the electrons are localized or itinerant, and therefore a determination of the magnetic form factor does not address that question; itinerancy is determined by whether or not the band crosses the Fermi surface. For the cuprates the single *d*-electron hole has $e_g$ symmetry, $x^2$-$y^2$, which is quite anisotropic and so makes the magnetic form factor anisotropic [5]. Moreover, the in-plane nearest-neighbor spins are aligned antiparallel, and therefore any net spin transferred onto the intervening oxygen ion cancels to first order, so that the effects of bonding are difficult to detect in magnetic form factor measurements. For $K_2IrCl_6$, on the other hand, the single electron occupies a linear combination of $t_{2g}$ orbitals, which again yields a quite anisotropic form factor [6]. It also gives rise to a non-collinear (atomic) spin density and separation of the charge and spin degrees of freedom. This latter property is amplified by the bonding to the Cl ions, where charge is transferred to all six Cl ions in the $IrCl_6$ complex but spin is transferred only to the two Cl ions that reside along the spin direction. This spin transfer onto the Cl is not cancelled by neighboring spins since the Cl are not shared, rendering the overall form factor highly anisotropic. This highly anisotropic behavior contrasts with the present iron-based superconductors, where band theory shows that all five $d$-bands are occupied and cross the Fermi energy. Therefore the magnetic electrons are itinerant in character, with a multi-orbital description that is expected to yield a magnetic form factor that is much closer to isotropic (as observed). Note in particular that if the $d$-bands have equal occupancies then the form factor has spherical symmetry. In the iron-based systems, however, any spin transferred to the As ion along the $b$-direction does not cancel since neighboring spins are parallel, making it easier to see these bonding effects. We also note that the in-plane exchange interactions are dramatically different along the $a$ versus the $b$ axis [14-16], which should be related to the anisotropic spin density distribution observed here.

## IV. Structural Phase Transition

In the previous investigation of the structural phase transition, which breaks the high temperature tetragonal symmetry in going to the low temperature orthorhombic structure, the in-plane structure was characterized in detail [7]. The structural transition occurs rather abruptly, with the long range antiferromagnetic developing at the same temperature [7,17]. The shift in the diffraction peaks related to the structural distortion in the $a$-$b$ plane were found to be symmetric, in that one crystal axis ($a$) increased while the other ($b$) decreased by the same amount. More recent studies in other systems [18] have found that the splitting can occur asymmetrically, concomitant with an abrupt change in the $c$-axis as well. We therefore carried out high resolution measurements of the (0,0,4) structural Bragg reflection to investigate whether there is any significant change in the $c$-axis for $SrFe_2As_2$ in going through the phase transition. Figure 3 shows radial (θ:2θ) scans through this reflection, which provide a direct determination of the $c$-axis lattice parameter. The horizontal (black) line indicates the temperature of the structural phase transition. The intensity of the reflection is indicated by the shading, and the inset shows a representative scan. There is some modest change in the intensity of the reflection, but we clearly see that the *c*-axis lattice parameter remains constant through the transition, in contrast to what is seen in Ca(Fe-Ni)$_2$As$_2$ [18].

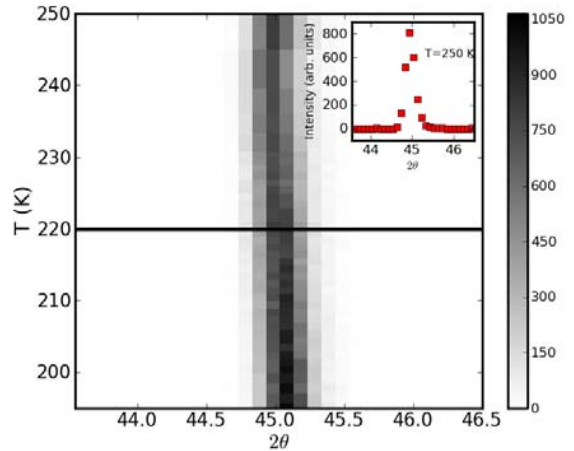

Figure 3. (color online) Intensity of the (0,0,4) Bragg peak, as measured in a radial (θ:2θ) scan, as a function temperature. The black horizontal line represents the temperature of the tetragonal-orthorhombic transition at ~220 K. No change in the $c$-axis lattice parameter is detected. The inset shows a representative scan. In most cases, error bars are smaller than the marker size.

**V. Conclusions**

In summary, we have measured the magnetic form factor in SrFe$_2$As$_2$, and found it to be approximately isotropic and in reasonable agreement with the form factor of metallic Fe. This behavior is consistent with electron occupancy of all five *d*-orbitals as expected from band theory calculations. Both Fourier inversion of the data and maximum entropy reconstructions suggest an elongation of the moment densities in the direction of As atoms, indicative of Fe-As bonding. We have also investigated the behavior of the *c*-axis through the structural phase transition, and found no significant anomaly as a function of temperature. This structural behavior differs from that observed in Ca(Fe-Ni)$_2$As$_2$, which exhibits an asymmetry in the in-plane distortion and an abrupt *c*-axis anomaly.

**Acknowledgments**
The work at UT/ORNL is supported by the US DOE, BES, through DOE DE-FG02-05ER46202, and in part by the US DOE, Division of Scientific User Facilities. The work at IOP, CAS is supported by CAS and 973 Program (2010CB833102).